\documentclass[12pt]{article}
\usepackage{epsfig}

\def\beq{\begin{equation}}
\def\eeq#1{\label{#1}\end{equation}}
\def\eeqn{\end{equation}}

\def\beqa{\begin{eqnarray}}
\def\eeqa#1{\label{#1}\end{eqnarray}}
\def\eeqan{\end{eqnarray}}

\let\bar=\overbar

\def\Dslash{\not{\hbox{\kern-4pt $D$}}}
\def\dslash{\not{\hbox{\kern-2pt $\del$}}}

\def\msb{{\bar{\ssstyle M \kern -1pt S}}}

\def\Title#1{\begin{center} {\Large {\bf #1} } \end{center}}

\begin{document}

\Title{Confinement of Color and Geometry}

\bigskip\bigskip

%+\addtocontents{toc}{{\it D. Reggiano}}
%+\label{ReggianoStart}

\begin{raggedright}  

{\it Adriano Di Giacomo\index{Di Giacomo,A.}\\
Department of Physics \\Pisa University and INFN Sezione di Pisa\\
3 Largo B. Pontecorvo
I-56127, ITALY}
\bigskip\bigskip
\end{raggedright}

\section{Confinement and Symmetry}

Quarks are not observed as free particles neither in Nature, nor as a product of high energy reactions. This phenomenon is known as Confinement of Color.

To quantify it one can compare the upper limits put by experiment on the observation of free quarks to the expectations obtained from  $QCD$ assuming that they can exist as  ordinary particles.

The upper experimental limit on the relative abundance of quarks with respect to protons, ${n_q\over n_p    }$ is \cite{PDG}
\begin{equation}
{{n_q } \over { n_p    }} \le 10^{-27}
\end{equation}
to be compared to the expectation in the Standard Cosmological Model \cite{Okun} ${n_q\over n_p    }\approx 10^{-12}$.

The upper limit to the cross section for  inclusive  production of a quark or anti-quark in $p+p$ collisions at the energy of the CERN SPS is \cite{PDG}
\begin{equation}
\sigma _q  \equiv \sigma (p+p \rightarrow q[\bar q] + X) \le 10^{-40} cm^2
\end{equation}
If quarks could exist as free particles the expectation for ${\sigma_q}$ would be a sizable fraction of
 the total cross section $\sigma_T$ , or, at the energy of the experiment, $\sigma_q \approx 10^{-25} cm^2$.
 
 The inhibition factor is smaller than $10^{-15}$ in both cases.
 
 The only natural explanation of this small number is then that both $n_q$ and $\sigma_q$ are strictly $zero$ as a consequence of a symmetry, so that confinement is an absolute property and the deconfining transition  a change of symmetry, i.e. an order-disorder transition.
 
 An order-disorder transition can not be a cross-over.
 
 In the phase diagram of $QCD$ as determined numerically on the Lattice there exist ranges of values of the parameters in which no discontinuity is observed within the numerical precision and within the limits of the spatial volume of the sample. The usual lore is to say that the transition is in fact a crossover\cite{owe}. Of course a numerical statement that all observables are continuous with all  derivatives in the
 infinite volume limit is impossible in principle. The popular statement has to be understood in the sense
 that the present data are compatible with a continuous transition. It could well be, however, that the transition is 
 a  weak first order, and that  the corresponding growth of, say, the specific heat with the volume only
 becomes visible at larger volumes than those which have been attained in the simulations\cite{ddp}\cite{cddp}.  The existence of a tri-critical point \cite{SS} as the end point of a line of first order transitions at non-zero baryon chemical potential would be an unambiguous indication that there exist
 cross-overs, and such would be the statement that the chiral transition at zero quark mass is second order. The tricritical point has not been found up to now , neither on the lattice nor in heavy ion collisions.
 The scaling laws in the neighborhood of the chiral point are not compatible with second order $O(4)$
 with present lattices. Everything is compatible with first order, weak enough, however, so that the growth with the volume of the susceptibilities is not visible at present volumes and lattice spacings\cite{ddp} \cite{cddp}.
 The issue is open: an order disorder transition and an explanation in terms of symmetry is not excluded. The real theoretical question is : what symmetry, if any?

\section{What symmetry?}

 The most known symmetries are the Noether's symmetries of the Lagrangean.
 
   In  $QCD$ there  is exact invariance  under $SU(3)$ gauge transformations, both in 
  the deconfined phase  and in the confined phase .    This means no change at deconfinement, or that  color symmetry can not be the symmetry we are looking for.
  
   A flavor symmetry exists at zero quark masses ($m_q=0$), the chiral symmetry. However it is not a symmetry in the realistic case of non zero quark masses. Moreover an indication that it is not the symmetry relevant to confinement comes from the studies of a variant of $QCD$ in which quarks belong to the adjoint representation of the color group. There, contrary to ordinary $QCD$ where quarks belong to  the fundamental representation, deconfinement and chiral transition do not coincide, but occur at different temperatures,
   clearly indicating that the degrees of freedom relevant to deconfinement  are different from those relevant to the chiral transition\cite{K},\cite{CDDLP}.
   
   In the quenched theory (no quarks) the Lagrangean is blind to the centre of the group , $Z_3$, which is then a symmetry. The order parameter is the Polyakov loop $\langle L \rangle$. The symmetry is spontaneously broken in the confined phase $T\le T_c$ and is restored above $T_c$ . However quarks do exist in Nature and therefore $Z_3$ can not be the symmetry we are looking for.
   
  An alternative is to look for a dual symmetry. 
   
   The idea is realized in a number of systems in statistical mechanics. These systems admit excitations with non trivial boundary conditions ( homotopy) and the 
   partition function can be expressed either in terms of the original local fields, or in terms of the dual excitations. An example is the $2d$ ising model\cite{KC}, whose dual excitations are kinks,  or the $3-d$ $(x-y)$ model, whose dual excitations are vortices \cite{DDPT} and the dual symmetry is the conservation of the number of vortices minus that of anti-vortices.
   
   In gauge theories dual excitations correspond to a violation of Bianchi identities and their nature is dictated by geometry, namely by the number of space dimensions. The conserved dual quantities are indeed related to the homotopy, i.e. to the mapping of the surface at spatial infinity onto a subgroup of the gauge group. 
   
   In $(2+1)$ dimensions the surface is a circle $S_1$, the homotopy group is $\Pi_1$, the excitations are vortices and the conserved quantity is the number of vortices minus the number of anti-vortices\cite{'tH1}.
   
   In $(3+1)$ dimensions the surface is the spherical surface in $3d$ space
     $S_2$, the homotopy group is $\Pi_2$ , the excitations  are monopoles and the magnetic charge the conserved quantity\cite{FM}\cite{DP}\cite{SW}.
     
   The only possible symmetry to describe confinement of color is then the conservation of magnetic charge .
   
   In the supersymmetric case of Ref,\cite{SW} electric-magnetic duality is proved explicitely. In the realistic case of ordinary $QCD$ it can only be a guess to be eventually demonstrated by numerical simulations on the lattice{\cite{cc}\cite{fqcd}.
   The origin of confinement goes then back to the old idea of Ref.'s \cite{m}, \cite{'thP} : below $T_c$ magnetic symmetry is Higgs-broken, the system is a dual superconductor generating a linear growing
   force between color charges at large distance by chromo-electric Abrikosov flux tubes. Above $T_c$, instead, magnetic symmetry of the vacuum is restored and confinement disappears. 
   
   Implementing correctly this idea  is far from trivial, as we shall see  below. 
   
   \section{The geometry of Monopoles}
   
   The prototype  monopole in gauge theories is the soliton classical solution of Ref.'s \cite{'tH}, \cite{Pol} of the $SU(2)$ Higgs model, with the Higgs field in the adjoint representation.
   
   In the so-called "hedgehog" gauge the Higgs field of this solution behaves at  distances large compared to the scale of the system, as 
   \begin{equation}
   \phi^i  \approx _{r\to \infty} \frac {r^i} {r}
   \end{equation}
   The direction of the Higgs field in color space coincides with the direction of the position vector $\vec r$.
   Eq.(3) describes a non trivial mapping of the sphere at infinity $S_2$ onto the group $SU(2)/U(1)$.
   
   At large distances the field strength of the solution behaves as follows
\begin{eqnarray}
   \vec E ^i&  \equiv &F_{0i}  =0\\
   \vec H^i &  \equiv & \frac {1} {2} \epsilon{ijk} F_{jk}= \frac {1} {g} \frac {\vec r} {4 \pi r^3}
   \end{eqnarray}
   A gauge invariant tensor $ F_{\mu \nu}$ can be defined, the 't Hooft tensor \cite{'tH},
   \begin{equation}
   F_{\mu \nu} = Tr(\phi G_{\mu \nu}) - \frac {i} {g} Tr( \phi [D_{\mu} \phi, D_{\nu} \phi])
   \end{equation}
   which coincides with the abelian field strength $F^3_{\mu \nu} = \partial _{\mu} A^{3}_{\nu}-\partial _{\nu} A^3_{\mu}$ in the unitary gauge in which the Higgs field is oriented along the third axis
   $\phi ^i = |\vec \phi| \delta^i_{3}$.
   
   For the soliton of Ref.'s \cite{'tH}, \cite{Pol} the 'tHooft tensor can be explicitly computed and is nothing but the field of a Dirac monopole with magnetic charge $\frac {2} {g}$ at all distances, irrespective of the specific value of the parameters of the model.
   \begin{eqnarray}
   \vec e &  =&0 \\
   \vec  h  &  =& {1\over  g}  { {\vec r} \over  {4\pi r^3} }+ Dirac - string
   \end{eqnarray}
    Here $\vec e$ and $\vec h$ are the electric and magnetic fields of the field strength $F_{\mu \nu}$.
    The solution violates abelian Bianchi identities, in a formulation in which the string is invisible,  since
    \begin{equation}
    \vec \nabla \vec h = {1 \over g} \delta^3(\vec x) \neq 0
    \end{equation}
    In a more formal way, if we define the dual of the 'tHooft tensor $ F_{\mu \nu}^* = \frac {1} {2} \epsilon_{\mu \nu \rho \sigma} F_{\rho \sigma}$,
    \begin{equation} 
    \partial _{\mu} F^*_{\mu \nu} = j_{\nu} \neq 0
    \end{equation}
    Due to the antisymmetry of  $F^*_{\mu \nu}$ the magnetic current  $j_{\nu}$ is conserved.
    \begin{equation}
    \partial _{\nu} j_{\nu} =0
    \end{equation}
    The conservation law Eq.(11) is the dual symmetry.
    
    It can be proved that Eq.(10) is the gauge invariant content of the non-abelian Bianchi identities\cite{bdlp} (See section 4 below)
    \begin{equation}
    D_{\mu} G^*_{\mu \nu} = J_{\nu}
    \end{equation}
    The magnetic current of Eq.(12) can be defined also in absence of Higgs breaking as well as 
    for a theory with no Higgs field at all. 
     
     Also the gauge group needs not be $SU(2)$ but can be generic\cite{DLP}. 
     
     The existence of a monopole requires a mapping of the sphere at spatial infinity onto a $SU(2)$ subgroup of the gauge group.
     In any compact gauge Lie group of rank $r$ with algebra spanned by the generators $H_i$ $(i = 1,   r)$ , $E_{\pm \vec \alpha}$,
     there exists an $SU(2)$ subgroup for each root $\vec \alpha$.
     
     Indeed the commutation relations , in the usual notation, read as follows
    \begin{eqnarray}
     [ {H_i} , {H_j} ]  & =& 0  \\
   {[} {H_i} , {E_{\pm \vec \alpha}} {]} & =&\pm \alpha _i  E_{\pm \vec \alpha} \\
    {[} E_{\vec \alpha} , E_{- \vec \alpha}{]} & =& \vec \alpha.\vec H  \\
    {[}E_{\vec \alpha},E_{ \vec \beta} {]}  &= &N_{\vec \alpha \vec \beta} E_{\vec \alpha+\vec \beta}
  \end{eqnarray}
  To each root $\vec \alpha$ an $SU(2)$ algebra can be associated 
  \begin{eqnarray}
   T^{\vec \alpha}_{\pm} &=& \sqrt{ { {2\over {(\vec \alpha \vec \alpha)}}}} E_{\pm \alpha}\\ T^{\vec \alpha}_{3}&=& {{\vec \alpha \vec H} \over {(\vec \alpha \vec \alpha)}} 
   \end{eqnarray} 
  
  A special role \cite{DLP} is played by the simple roots, defined as positive roots which are not equal to the sum of other positive roots [See e.g. \cite{la}]. They are in one-to-one correspondence with the little circles in the Dynkin diagram of the group: there are as many of them as the rank $r$ of the group.
  
  To each simple root $\vec \alpha^i$ an $r$-dimensional vector $\vec c^i$ can be associated such that
  \begin{equation}
  \vec \alpha^i . \vec c^j  = \delta_{ij}
  \end{equation} 
  The vectors $\vec c^i$ are called fundamental weights. 
  
  If one thinks of a coupling to a Higgs field and of a Higgs breaking  giving monopole solutions,
  it can be shown that Higgs fields $\phi^i$, which in the unitary gauge are proportional to
  $\mu ^i= \vec c^I.\vec H$, identify 'tHooft-Polyakov monopoles embedded in the group,
  and identify the corresponding 'tHooft tensors $F^i_{\mu \nu}$.
  In the unitary gauge 
  \begin{equation}
  F^i_{\mu \nu} =  \partial _{\mu} A^{i3}_{\nu} - \partial_{\nu} A^{i3}_{\mu}
  \end{equation}
  $A^{i3}_{\mu}$ is the component of the field along the third axis of the $SU(2)$ subgroup Eq's.(17),(18).
  $\mu^i$ is the analog of $\sigma^3$ in the $S(2)$ case.
 The  $F^i_{\mu \nu} $'s  can be given a gauge invariant form, analogous to that of Eq.(6) \cite{DLP}, and define $r$ independent magnetic charges, corresponding to the violation of Bianchi identities.
  \begin{equation}
  \partial_{\mu} F^{i*}_{\mu \nu} = j^i_{\nu} 
  \end{equation}
  The conservation laws $(i= 1, ..r)$
  \begin{equation}
  \partial_{\nu} j^i_{\nu} =0
  \end{equation}
  are the dual symmetries and their Higgs-breaking is responsible for confinement. Their restoration corresponds to the deconfining transition.
  
  The general form of the t'Hooft tensor is \cite{DLP}
  \begin{eqnarray}
  F^i_{\mu \nu}& = &Tr(\phi^i G_{\mu \nu}) - {i \over g} \Sigma_I {1\over {\lambda^i _I} }Tr(\phi^i {[}D_{\mu}\phi^i, D_{\nu} \phi^i {]}) \\ &+&{i \over g} \Sigma_{IJ} {1\over {\lambda^i_I \lambda ^i_J}} Tr(\phi^i {[}{[}\phi^i,D_{\mu}\phi{]}, {[}\phi^i, D_{\nu}\phi^i{]}{]}) \nonumber \\ &-&{i \over g}\Sigma_{IJK} {1\over {\lambda^i _I \lambda ^i _J \lambda^i _K}}....\nonumber
  \end{eqnarray}
    Here $\phi^i$ is an operator in the orbit of the fundamental weight $\mu^i$ i.e.
    \begin{equation}
    \phi^i = U^{\dagger}(x) \vec c^i. \vec H U(x)
    \end{equation}
    with U(x) any element of the gauge group.
    
    $\lambda^i_I$ are the non zero values which the quantity $(\vec c^i .\vec  \alpha)^2$ can assume on the set of  all the roots $\vec \alpha$  . Since all positive roots are sums of simple roots, as a consequence of 
    Eq.(19) $|(\vec c^i .\vec  \alpha)|$ counts the number of times that a simple root appears in the given root, and is an integer.  The sums in Eq.(23) run on the possible values of $\lambda^i_I$, each of them taken once. The coefficients of the expansion Eq.(23) are uniquely determined by the simple root considered (index $i$) and by the Lie algebra of the group\cite{DLP}.
    
    As an example in $SU(N)$ for arbitrary $N$ each root contains any simple root at most once,$\lambda^i_I$ assumes only one non zero value, namely $1$, only the first term of Eq.(19) survives
    and the expression for the 'tHooft tensor reduces to that of $SU(2)$ Eq.(6).
    
    \section{Non abelian Bianchi identities\cite{bdlp}}
    The non abelian version of the Bianchi identities reads
    \begin{equation}
    D_{\mu} G^*_{\mu \nu} = J_{\nu}
    \end{equation}
    The Bianchi identities demand $J_{\nu} = 0$ : a non-zero $J_{\nu}$ means violation of them.
    Here
    \begin{equation}
    G^*_{\mu \nu} \equiv {1\over 2} \epsilon_{\mu \nu \rho \sigma} G_{\rho \sigma}
    \end{equation}
    is the dual tensor to the field strength $G_{\mu \nu}$. It is easily shown that
    \begin{equation}
    D_{\nu} J_{\nu} =0
    \end{equation}
    The magnetic current is covariantly conserved. This is not a conservation law in the usual sense.
    On the other hand Eq.(25) is not gauge invariant. We can, however , isolate the gauge invariant  content of it .  To do that we can e.g. diagonalize the left-hand side of Eq.(25) and look for independent
    matrix elements: there are $r$ of them, $r$ being the rank of the group. This can be done by projecting 
    on the fundamental weights $\mu^i$ or on any element on their orbit $\phi^i$ [See Eq.(24)]. In formulae\cite{bdlp}
    \begin{equation}
    Tr(\phi^i D_{\mu}G^*_{\mu \nu}) = Tr(\phi^i J_{\nu})
    \end{equation}
    From the identity
    \begin{equation}
    \partial _{\mu} Tr(\phi^i G^*_{\mu \nu}) = Tr(D_{\mu} \phi^i G^*_{\mu \nu}) + Tr(\phi D_{\mu} G^*_{\mu \nu})
    \end{equation}
    inserting Eq.(25) we get
    \begin{equation}
    \partial _{\mu} Tr(\phi^i G^*_{\mu \nu}) - Tr(D_{\mu} \phi^i G^*_{\mu \nu}) =Tr(\phi^i J_{\nu})
    \end{equation}
    It can be also shown that identically \cite{bdlp}
    \begin{equation}
     \partial _{\mu} Tr(\phi^i G^*_{\mu \nu}) - Tr(D_{\mu} \phi^i G^*_{\mu \nu}) =\partial _{\mu} F^i_{\mu \nu}
     \end{equation}
     so that finally the gauge invariant content of the non abelian Bianchi identities is equivalent  to the $r$ abelian Bianchi identities for the different monopole species.
     All this is part of a research program aiming to give monopoles a fully gauge invariant formulation.

\end{document}